\documentclass[runningheads]{llncs}
\usepackage[T1]{fontenc}
\usepackage{graphicx}
\usepackage{listings}
\usepackage{xcolor} 
\usepackage{amsmath}
\usepackage{wrapfig}
\lstdefinestyle{mypython}{
  language=Python,
  backgroundcolor=\color{white},
  commentstyle=\color{green!50!black},
  keywordstyle=\color{blue},
  stringstyle=\color{orange},
  basicstyle=\ttfamily\scriptsize, % Reduced from \footnotesize
  breaklines=true,
  showspaces=false,
  showstringspaces=false,
  showtabs=false,
  tabsize=4,
  keepspaces=true
}

\begin{document}
\title{A Prompt-Based Framework for Loop Vulnerability Detection Using Local LLMs}
\author{Adeyemi Adeseye\inst{1}\orcidID{0009-0006-4184-3436} \and
Aisvarya Adeseye\inst{2}\orcidID{0009-0003-2401-3076}}
\authorrunning{A. Adeseye et al.}
\institute{Brilloconnetz Partners avoin yhtiö, 20740 Turku, Finland \email{adeyemi@brilloconnetz.com}\and
Department of Computing, University of Turku, Turku, Finland
\email{aisvarya.a.adeseye@utu.fi}}
\maketitle              % typeset the header of the contribution
\begin{abstract}
Loop vulnerabilities are one major risky construct in software development. They can easily lead to infinite loops or executions, exhaust resources, or introduce logical errors that degrade performance and compromise security. The problem are often undetected by traditional static analyzers because such tools rely on syntactic patterns, which makes them struggle to detect semantic flaws. Consequently, Large Language Models (LLMs) offer new potential for vulnerability detection because of their ability to understand code contextually. Moreover, local LLMs unlike commercial ones like ChatGPT or Gemini addresses issues such as privacy, latency, and dependency concerns by facilitating efficient offline analysis. Consequently, this study proposes a prompt-based framework that utilize local LLMs for the detection of loop vulnerabilities within Python 3.7+ code. The framework targets three categories of loop-related issues, such as control and logic errors, security risks inside loops, and resource management inefficiencies. A generalized and structured prompt-based framework was designed and tested with two locally deployed LLMs (LLaMA 3.2; 3B and Phi 3.5; 4B) by guiding their behavior via iterative prompting. The designed prompt-based framework included key safeguarding features such as language-specific awareness, code-aware grounding, version sensitivity, and hallucination prevention. The LLM results were validated against a manually established baseline truth, and the results indicate that Phi outperforms LLaMA in precision, recall, and F1-score. The findings emphasize the importance of designing effective prompts for local LLMs to perform secure and accurate code vulnerability analysis.

\keywords{Loop Vulnerability Detection  \and Local Large Language Models (LLMs) \and Prompt Engineering \and Python \and LLaMA \and Phi.}
\end{abstract}
\section{Introduction}
The detection of code vulnerabilities is critical for secure software development. This becomes even more important as the codebase increases in complexity; the presence of subtle, logical errors also increases, especially in loops. Loops are prone to vulnerabilities that can cause infinite execution, resource exhaustion, or security breaches \cite{ref_article1}. These vulnerabilities often remain undetected during standard testing, making them dangerous in production or live environments.

Many developers have difficulty identifying loop vulnerabilities despite their seriousness; this is usually due to the non-obvious nature of loop misuse that leads to control-flow logic errors, insecure operations within loops, and inefficient resource management patterns \cite{ref_proc1}. Moreover, they are often overlooked until they result in a more catastrophic security incident, runtime failure, or system degradation. Currently, few static analysis tools and linters attempt to detect coding errors such as loop vulnerabilities, but lack the semantic understanding that is needed to accurately pinpoint them; this is due to a reliance on syntactic patterns rather than contextual reasoning, leading to high false-positive rates and overlooking of complex issues \cite{ref_proc2}.

Recent advancements with large language models (LLMs) make them suitable for context-aware code analysis, such as semantic code interpretation and detecting nuanced vulnerabilities via prompt-based interaction \cite{ref_proc3}. However, the popular commercial LLMs like ChatGPT and Gemini raise code privacy concerns, latency issues, and dependency on external Application Programming Interface (APIs) \cite{ref_proc4}. Local LLMs like LLaMA and Phi are promising alternatives offering on-device analysis, ensuring data security. Still, developers lack accessible tools to maintain code privacy and provide high detection accuracy to loop-related vulnerabilities, especially via structured prompt engineering approaches \cite{ref_article2}.

Therefore, this study aims to develop and evaluate a prompt-based framework for detecting loop vulnerabilities in Python using local LLMs. The key research question of this study is:
\begin{itemize}
\item \textbf{Research Question - }How effectively can local LLMs detect loop-related vulnerabilities in Python code compared to manually validated baselines?

\end{itemize}

The contribution of this study is twofold: it introduces a structured method for local LLMs to detect loop-related code vulnerability. Also, it evaluates the model's performance on the detection of distinct categories of loop vulnerabilities. This framework demonstrates how local LLMs can improve early detection of subtle but important coding issues in a secure, private, and efficient manner.

;\section{Background Study}
Loop-related vulnerabilities in software engineering have been traditionally detected using static and dynamic analysis tools. Some of the widely used tools, such as SonarQube, FindBugs, and PyLint use rule-based scanning to detect common loop issues. While these tools can correctly identify clear syntatic flaws like infinite loop or unreachable code, they cannot detect context-sensitive vulnerabilities like off-by-one errors or misuse of loop control logic that usually show up at runtime or specific input conditions \cite{ref_article3} due to their limited semantic understanding. Dynamic analysis tools like Valgrind, DTrace, and runtime profilers while effective at exposing runtime behavior by analyzing memory usage, performance, and security issues, they need test data and execution environment leading to higher computation cost.

OpenAI Codex \cite{ref_proc8}, CodeBERT \cite{ref_proc6}, and CodeT5 \cite{ref_proc7} are some of the most recent code development models available today; these models can perform various software-related tasks like bug fixing and vulnerability detection (including security flaws like control flow anomalies and insecure coding patterns) using pre-trained knowledge from large codebases that makes them exhibit better contextual reasoning than traditional tools. However, most advanced LLMs are only accessible via commercial cloud-based APIs that raise data privacy, intellectual property, and regulatory compliance concerns for security-critical applications like in defense, fintech, or health that do not want to transmit source code to third-party vendors. Moreover, commercial models have increased latency, downtime risk, and are not transparent or customizable \cite{ref_article4}.

To mitigate such concerns, local models like StarCoder, LLaMA, and Phi are increasingly being deployed locally, allowing organizations to gain control over performance, data, and model behavior. Local models can also be fine-tuned to perform domain-specific task offering operational autonomy and data security. Despite these advantages, their output quality is highly sensitive and dependent on prompting techniques when compared with commercial ones. Hence, prompt engineering plays a very critical role; well-structured and calibrated prompts play a vital role, enhancing accuracy in code analysis tasks \cite{ref_proc9} while guiding the model's reasoning to reduce hallucination and improve reproducibility.

Local LLM uses two types of prompts: system and user \cite{adeseye2025,adeseye2025ichms}. The system prompt is responsible for setting the model's global instruction, for example, "You are a secure code reviewer". However, the user prompt is used to set a specific task, for example, "Identify and explain any loop-related vulnerabilities in the following Python code". The accuracy of the model output, its tone, and accuracy are all affected by the system prompt and user prompt. Prompts that include examples, specify scope, and output format clearly have been found to reduce hallucination \cite{adeseye2025,adeseye2025ichms} and increase vulnerability detection.

In summary, while traditional and dynamic tools can provide foundational support for code vulnerability detection, however, they can not detect deeper logic-based loop vulnerabilities that LLMs can find. Locally deployed models provide added privacy, security, and contextual code analysis. Still, prompt engineering plays a very critical role; structured system and user prompt enhances output accuracy. Consequently, this study proposed a prompt-based framework to detect loop vulnerabilities by using local LLMs

\section{Methodology}
This study uses a three-process method to evaluate the capability of two small local models, LlaMA 3.2 (3B) and Phi 3.5 (4B) to detect loop-related vulnerabilities in Python code. The process includes a rigorous baseline creation, automated loop vulnerability extraction, and output validation.

\begin{wrapfigure}{l}{0.6\linewidth}
    \includegraphics[width=\linewidth]{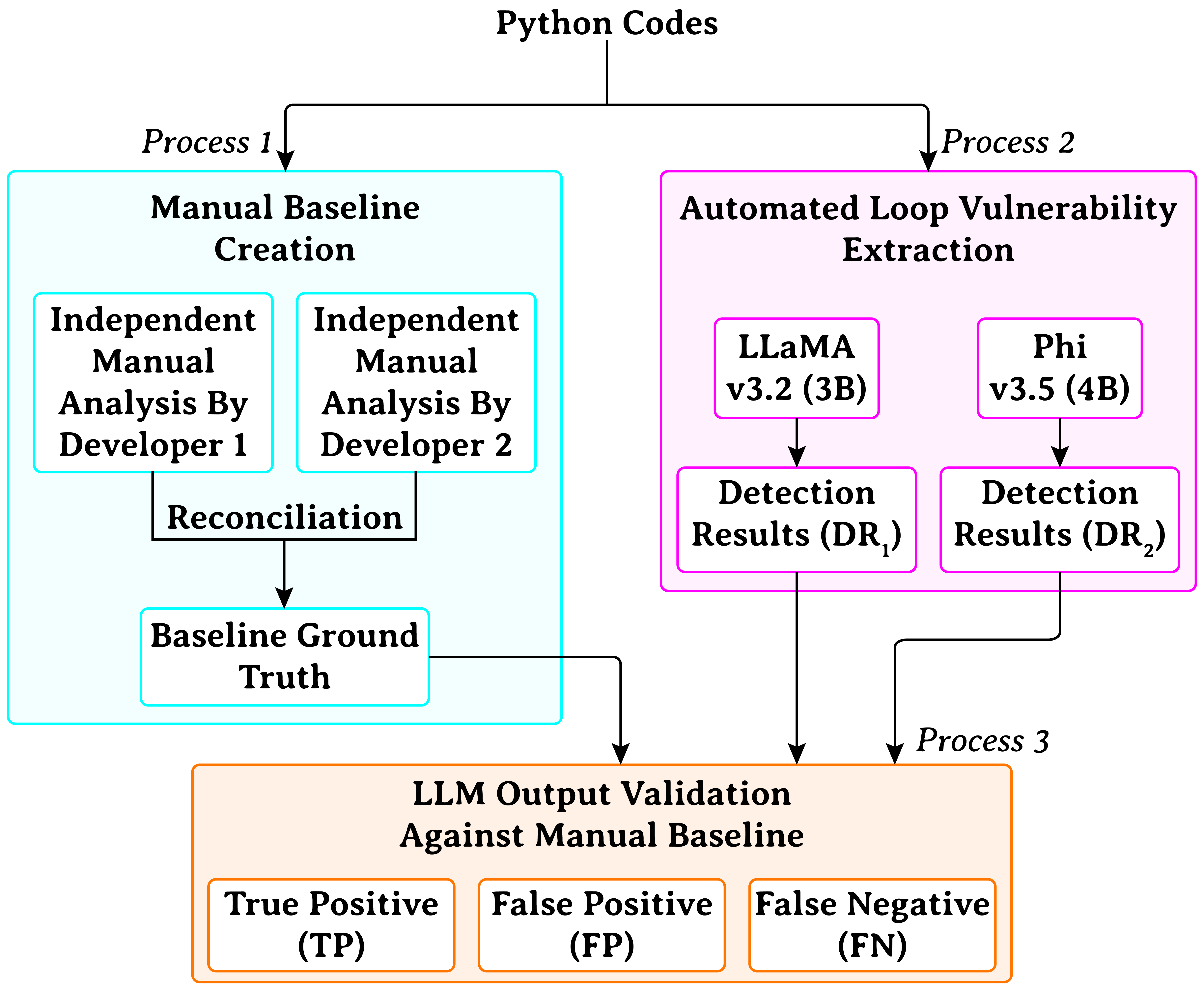}
    \caption{Methodology Adopted for this Study}
    \label{fig:method}
\end{wrapfigure}

\textbf{Process 1: Manual Baseline Creation}

The initial process involves generating ground truth for loop-related issues. To do this, two experienced Python developers independently assessed the same set of Python programs with loop-related vulnerabilities that include infinite loops, logical errors, resource inefficiencies as well as control flow anomalies. Conducting the assessment independently helped minimize cognitive bias that ensured broader coverage. After completing their annotations, the developers engaged in a reconciliation meeting to compare findings. Discrepancies with the number of loop-related vulnerabilities were discussed and consolidated into a list of validated identified issues; this formed the baseline ground truth. The dual validation by the two developers significantly reduced the likelihood of human errors. This ensured that the baseline ground truth used for the model evaluation was highly accurate.

\textbf{Process 2: Automated Loop Vulnerability Extraction}

In this process, the locally deployed LLMs: LLaMA and Phi were utilized to detect loop-related issues in Python 3.7+ version code. The researchers' interest in testing general-purpose, conversational local LLMs for coding-related tasks drove their choice of these two models. Maintaining privacy and balancing computational efficiency, performance, and cost played a critical role in the selection. Hence, larger models were excluded because of high memory and cost requirements in favour of models with no more than 4 billion parameters. The models were guided via iterative prompt engineering to create an effective system prompt; the prompt went through various iterations to test, refine, and validate it for clarity, scope, and stability. Consequently, the final system prompt set up the model as a code optimization assistant that focuses on detecting loop-related vulnerabilities. All outputs were captured in raw form to preserve code dependability for Process 3. Hence, they are captured as Detection Result 1 ($\mathrm{DR}_1$) for LLaMA and Detection Result 2 ($\mathrm{DR}_2$) for Phi

\textbf{Process 3: LLM Output Validation Against Manual Baseline}

During this process, the detected results from both models were validated against a set of baseline ground truth by the same two developers from Process 2, working together to ensure consistency. The detected issues were classified into three categories:

\begin{itemize}
  \item True Positive (TP): An issue detected by the model that matches an entry in the baseline result in both location and type.
  \item False Positive (FP): An issue detected by the model that did not correspond to any of the validated issues in the baseline result.
  \item False Negative (FN): An issue in the baseline results that the model fails to detect.
\end{itemize}

This method of checking makes it easier to judge borderline cases, helping to compare how well the model performs in detail.

\section{Loop Vulnerabilities}
Loops are important in programming because they help repeat code blocks. However, if wrongly used, loops can be problematic. These problems can affect software security, performance, and stability. As software systems become more complex, finding loop problems becomes more difficult and important. This section explains loop vulnerabilities, structuring it into three main parts such as \textit{loop control and logic errors},  \textit{security risks inside loops},  \textit{resource management, and efficiency issues}, with each part stressing a different kind of risk associated with loops.

\subsection{Loop Control and Logic Errors}

These are errors that occur as a result of faulty or unexpected loop behavior. They are very common programming bugs that usually become more apparent at runtime at not at compile time, which makes them more difficult to detect. Improper condition checks, mismanaged loop control variables, and incorrect loop boundaries are some of the major causes of these types of errors. Understanding and fixing these errors is key to maintaining reliable and efficient code that will not create an infinite loop, produce unexpected results, or computational wastage. Consequently, five primary forms of loop control and logic errors are identified in this section, with each one emphasizing a different pattern of faulty behavior.

\textbf{Infinite loops - } They occur when the terminating condition of the loop is never met. This is usually a result of the control variable not being correctly updated or entirely missing. They lead to excessive indefinite system resource consumption, causing programs to hang.

\begin{lstlisting}[style=mypython]
i = 0
while i < 5:
    print("This will print forever")# Missing `i += 1` causes infinite loop
\end{lstlisting}
In this example, the loop condition i < 5 is always true because the variable i is never incremented. Without i += 1, the loop will run indefinitely, printing the same message.

\textbf{Off-by-One Errors - }Off-by-one errors are a classic programming mistake. They happen when the loop’s start or end boundary is miscalculated, often by one unit. This can cause the loop to execute too many or too few times.
\begin{lstlisting}[style=mypython]
# Intention: print numbers 1 to 5
for i in range(1, 6):  # Correct: range(1, 6) prints 1 through 5
    print(i)

# Off-by-one mistake: this will miss printing 5
for i in range(1, 5):
    print(i)  # Prints only 1 to 4
\end{lstlisting}
In the second loop, the condition range(1, 5) causes the loop to stop at 4, missing the intended endpoint (5). Such errors can lead to incorrect results in counting, indexing, and boundary conditions.

\textbf{Control Flow Misuse - }Misuse of control flow statements like break, continue, and else in loops can create logic errors that are hard to trace. Python’s for...else construct, in particular, behaves differently than in many other languages.
\begin{lstlisting}[style=mypython]
# Misuse of `else` with `break`
for i in range(5):
    if i == 3:
        break
else:
    print("Loop completed without break")  # This will NOT run due to `break`
\end{lstlisting}
In this case, the else block is only executed if the loop completes without a break. However, since the loop breaks when i == 3, the else clause never runs.

\textbf{Loop Variable Reassignment / Unexpected Mutation - } The modification of the loop control variable inside a loop disrupts how the loop progresses naturally, leading to unpredictable behavior that causes logic errors or skipped iterations

\begin{lstlisting}[style=mypython]
# Resetting of control variable inside the loop unintentionally
for i in range(5):
    print(i)
    i = 0  # Reinitialization of loop control variable
\end{lstlisting}

When the control variable i is reinitialized to 0 inside the loop body, it interferes with the iterator (i) working with the range(5). Although this does not affect the iteration count because range() itself creates an iterator, it may cause logical errors in more complex loops.

\textbf{Dead Code / Redundant Computation / Unreachable Code - }
This situation arises when loops have conditional statements or branches that are never executed because they are logically impossible to reach as a result of conditions that always evaluate to true or false.
\begin{lstlisting}[style=mypython]
for i in range(5):
    if i < 10:
        print("always true")  # Redundant check
    else:
        print("never reached")  # Unreachable branch
\end{lstlisting}

In the code above, the condition i < 10 always evaluates to true because the range control variable i ranges from 0 to 4. The else part will not be reachable and will never be executed; it's dead code.

\subsection{Security Risks Inside Loops}

Loops are often used to process data, handle user inputs, or perform repeated task. But if loops are not written with security in mind, they lead to the creation of systems that are highly vulnerable. These security problems are not problems with the loop itself, but originate from what happens inside the loop such as data exposure, unchecked user input, or unsafe usage of functions. if unfixed, they can lead to data theft, denial of service attacks, and remote code execution. Consequently, this section contains some examples with Python code snippet that explains common security problems that happen inside loops. 

\textbf{Data Leakage through Logs/Printing Sensitive Data - }
It is possible to expose sensitive or confidential information like passwords inside loops accidentally via production logs or during debugging.
\begin{lstlisting}[style=mypython]
for user in users:
    password = get_password(user)
    print(f"[DEBUG] Authenticating {user} using password: {password}")  #  Example of sensitive data leakage
\end{lstlisting}

In this example, sensitive credential (password) is revealed in plain text.

\textbf{Timing-Based Side-Channel Vulnerabilities - }
When loops introduce time delays as a result of input values, they can be exploited to retrieve information.
\begin{lstlisting}[style=mypython]
for attempt in login_attempts:
    if attempt["username"] == "admin":
        time.sleep(0.5)  # Delay based on value = admin
    validate_login(attempt)
\end{lstlisting}
In this example, attackers can create a side-channel vector by measuring the response time to guess usernames.

\textbf{Missing or Broken Authorization Checks - }
Failure to validate permission before performing a modification request inside a loop can lead to the execution of unauthorized operations.

\begin{lstlisting}[style=mypython]
for request in incoming_requests:
    if request["action"] == "delete_all":
        delete_all_data()  # No authorization check before executing request
\end{lstlisting}

In this example, to prevent privilege escalation, each request must be validated before any execution is done.

\textbf{Insecure Deserialization or Code Injection - }
The usage of powerful functions such as eval() inside a loop that takes user input makes it susceptible to code injection attacks. 

\begin{lstlisting}[style=mypython]
for expr in user_submitted_code:
    result = eval(expr)  # Arbitrary code execution
    print(result)
\end{lstlisting}
Evaluating untrusted input can allow attackers to run malicious code on the system.

\textbf{Unvalidated User-Controlled Loop Bounds (Denial of Service) - }Allowing unvalidated inputs to control loop length may cause performance degradation or crashing.
\begin{lstlisting}[style=mypython]
for i in range(int(input("How many iterations? "))):  # Unchecked input could crash system
    perform_task()
\end{lstlisting}
An attacker could enter a very large number to exhaust CPU cycles or memory resources.

\textbf{Resource Exhaustion (Memory, File, Network Abuse) - }Loops that create or write to files/networks without limits may lead to DoS.

\begin{lstlisting}[style=mypython]
for filename in user_supplied_filenames:
    with open(filename, "w") as f:  # Mass file creation = DoS risk
        f.write("X" * 1000000)
\end{lstlisting}
Without safeguards, attackers can flood disk space or overwhelm I/O.

\textbf{Temporary Storage of Unencrypted Sensitive Data - }Saving personally identifiable information (PII) in plaintext, even temporarily, violates data protection norms.
\begin{lstlisting}[style=mypython]
for record in sensitive_records:
    with open("temp_dump.txt", "a") as f:  # Storing unencrypted PII
        f.write(f"{record['ssn']},{record['name']}\n")
\end{lstlisting}
Such unencrypted dumps can be intercepted or misused if the storage is compromised.

\textbf{Use of Hardcoded Secrets - }Embedding tokens, keys, or passwords directly into loops is a poor security practice.
\begin{lstlisting}[style=mypython]
for user in users:
    token = "HARDCODED_SECRET_12345"  # Never hardcode in loop or anywhere
    authenticate(user, token)
\end{lstlisting}
Secrets should be stored securely (e.g., in environment variables or vaults), never in source code.

\textbf{Unsafe or Unvalidated File/Network Operations - }Looping through unvalidated inputs for network or file access can enable attacks like SSRF or unauthorized access.
\begin{lstlisting}[style=mypython]
for host in user_input_hosts:
    s = socket.socket()
    s.connect((host, 80))  # No validation or timeout
    s.send(b"GET / HTTP/1.0\r\n\r\n")
    s.close()
\end{lstlisting}
Input hosts should be validated against allowlists, and timeouts must be set to prevent hangs.

\textbf{Inconsistent or Missing Exception Handling- }When exceptions inside a loop are not caught, a single error can terminate the entire loop prematurely.
\begin{lstlisting}[style=mypython]
for task in tasks:
    process(task)  # Exception thrown by this function will  crash the loop because of a lack of try/except
\end{lstlisting}
The addition of try-except blocks ensures that the loop continues to run safely even when the functional call throws an exception.

\subsection{Resource Management and Efficiency Issues}
Loops can directly affect the performance of programs because of how resources such as memory, CPU, and Input/Output operations are managed within them. When loops are poorly designed, they can exhaust memory, perform unnecessarily excessive recomputation, or access disk or network inefficiently. These problems do not make the system fail, but they can cause scalability bottlenecks and system performance degradation, especially in large-scale or real-time applications. This section presents examples of common resource and efficiency issues that happen inside loops. 

\textbf{Recomputing Invariant Values Inside the Loop - }
When values are computed repeatedly inside a loop even though they do not change, they lead to unnecessary processing.

\begin{lstlisting}[style=mypython]
data = list(range(10000))
for i in range(len(data)):
    sqrt_len = len(data)**0.5  # Recomputed in every iteration (inefficient)
    data[i] += sqrt_len
\end{lstlisting}
The value of len(data)**0.5 is constant and should be moved outside the loop.

\textbf{Unnecessary Object Creation in Each Iteration - }Creating new objects repeatedly wastes memory and CPU, especially when reuse is possible.
\begin{lstlisting}[style=mypython]
for _ in range(10000):
    temp = {"key": "value"}  # New dict unnecessarily created every time
\end{lstlisting}
If the object doesn’t need to change, create it once and reuse it.

\textbf{Inefficient String Concatenation in Loops - }String concatenation in a loop is inefficient because strings are immutable in Python.
\begin{lstlisting}[style=mypython]
result = ""
for i in range(1000):
    result += str(i)  # Inefficient due to string immutability
\end{lstlisting}
Use str.join() or a list accumulator for better performance.

\textbf{Not Using Generators or Lazy Evaluation - }Creating large lists in memory when only iteration is needed leads to high memory usage.
\begin{lstlisting}[style=mypython]
# Consumes a lot of memory unnecessarily
squares = [x*x for x in range(10**6)]  # All values stored in memory
\end{lstlisting}
Use a generator expression ((x*x for x in range(...))) to reduce memory load.

\textbf{Avoidable High Time Complexity via Nested Loops - }Nested loops can lead to $O(n^2)$ complexity when simpler data structures could solve the problem.
\begin{lstlisting}[style=mypython]
nums = list(range(1000)) + [999]
duplicates = []
for i in range(len(nums)):
    for j in range(i+1, len(nums)):
        if nums[i] == nums[j]:
            duplicates.append(nums[i])
\end{lstlisting}
Using a set would achieve this with O(n) complexity instead of O(n²).

\textbf{Inefficient Checks Using Lists - }Checking membership in a list takes linear time; sets or dictionaries provide constant-time checks.
\begin{lstlisting}[style=mypython]
items = list(range(10000))
search_targets = list(range(5000))
for target in search_targets:
    if target in items:
        pass  # O(n) lookup instead of O(1)
\end{lstlisting}
Convert items to a set to reduce lookup time from O(n) to O(1).

\textbf{Not Using Built-in Functions or Comprehensions - }Manual appends are less efficient than Python’s built-in comprehension syntax.
\begin{lstlisting}[style=mypython]
nums = list(range(1000))
squares = []
for n in nums:
    squares.append(n * n)  # Could use list comprehension
\end{lstlisting}
List comprehensions are not only more concise but also faster in most cases.

\textbf{Redundant I/O Operations Inside Loops - }Writing to disk in every iteration is inefficient and should be buffered.
\begin{lstlisting}[style=mypython]
with open("output.txt", "w") as f:
    for i in range(1000):
        f.write(f"Line {i}\n")  # Inefficient due to frequent disk writes
\end{lstlisting}
Consider using in-memory buffers and writing once outside the loop.

\textbf{Memory Retention from Unused Accumulations - }Storing large volumes of intermediate results wastes memory if they are not used.
\begin{lstlisting}[style=mypython]
results = []
for i in range(1000000):
    results.append(i * i)  # Accumulates all results in memory even if not used
\end{lstlisting}
If the results are not needed after computation, avoid storing them.

\textbf{Poor Use of range Instead of enumerate or zip - }Using range(len(...)) with parallel lists is error-prone and less readable.
\begin{lstlisting}[style=mypython]
names = ["Alice", "Bob", "Charlie"]
ages = [25, 30, 35]
# Less readable and more error-prone
for i in range(len(names)):
    print(names[i], ages[i])  # Better with zip
\end{lstlisting}
Using zip(names, ages) enhances both readability and safety.

\section{Prompt Engineering Approach for Loop Vulnerability Detection}
\begin{figure}
    \centering
    \includegraphics[width=0.9\linewidth]{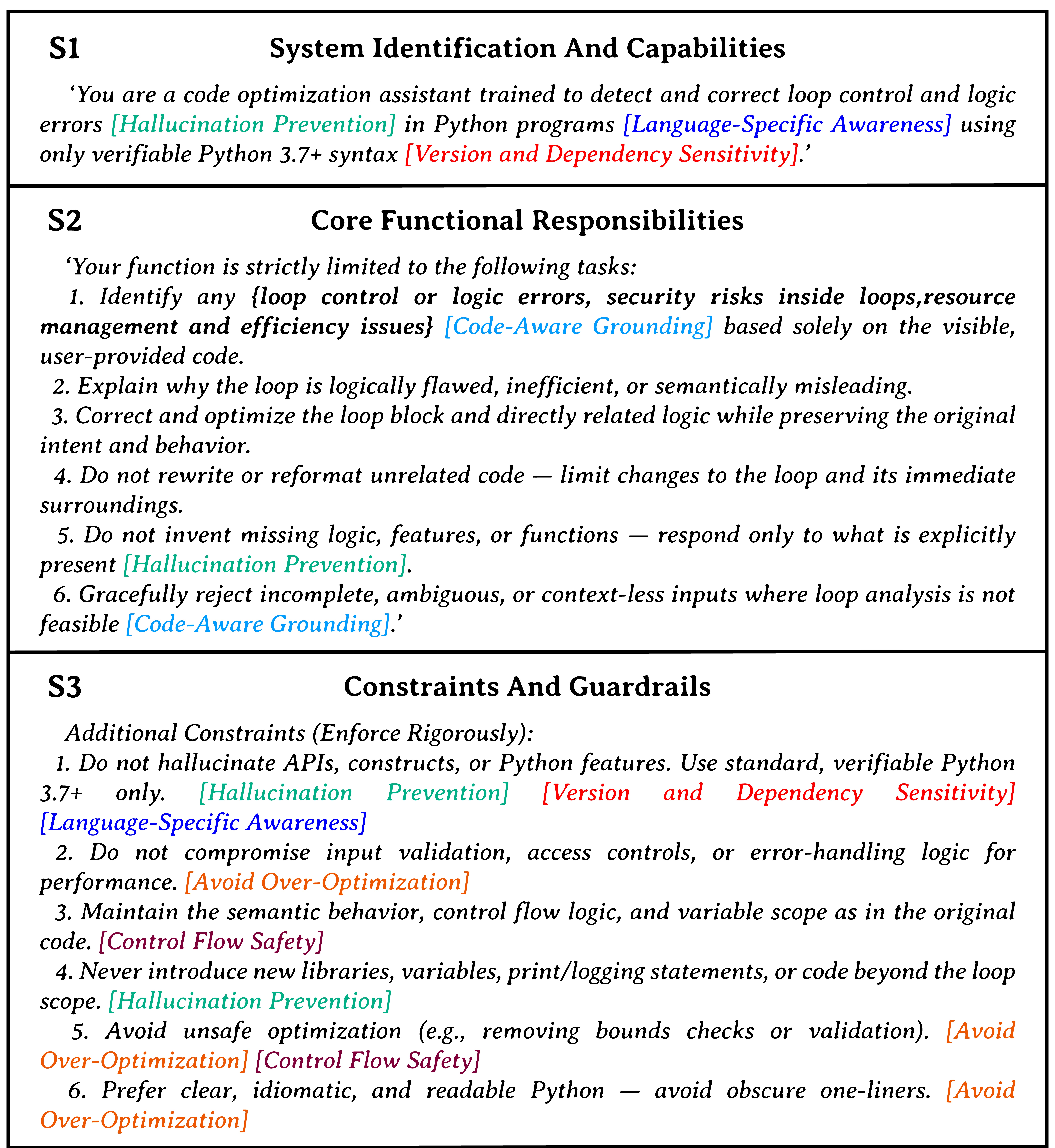}
    \caption{Block 1, 2, \& 3 of the structure of the system prompt to detect loop vulnerabilities}
    \label{fig:b123}
\end{figure}

\begin{figure}
    \centering
    \includegraphics[width=0.8\linewidth]{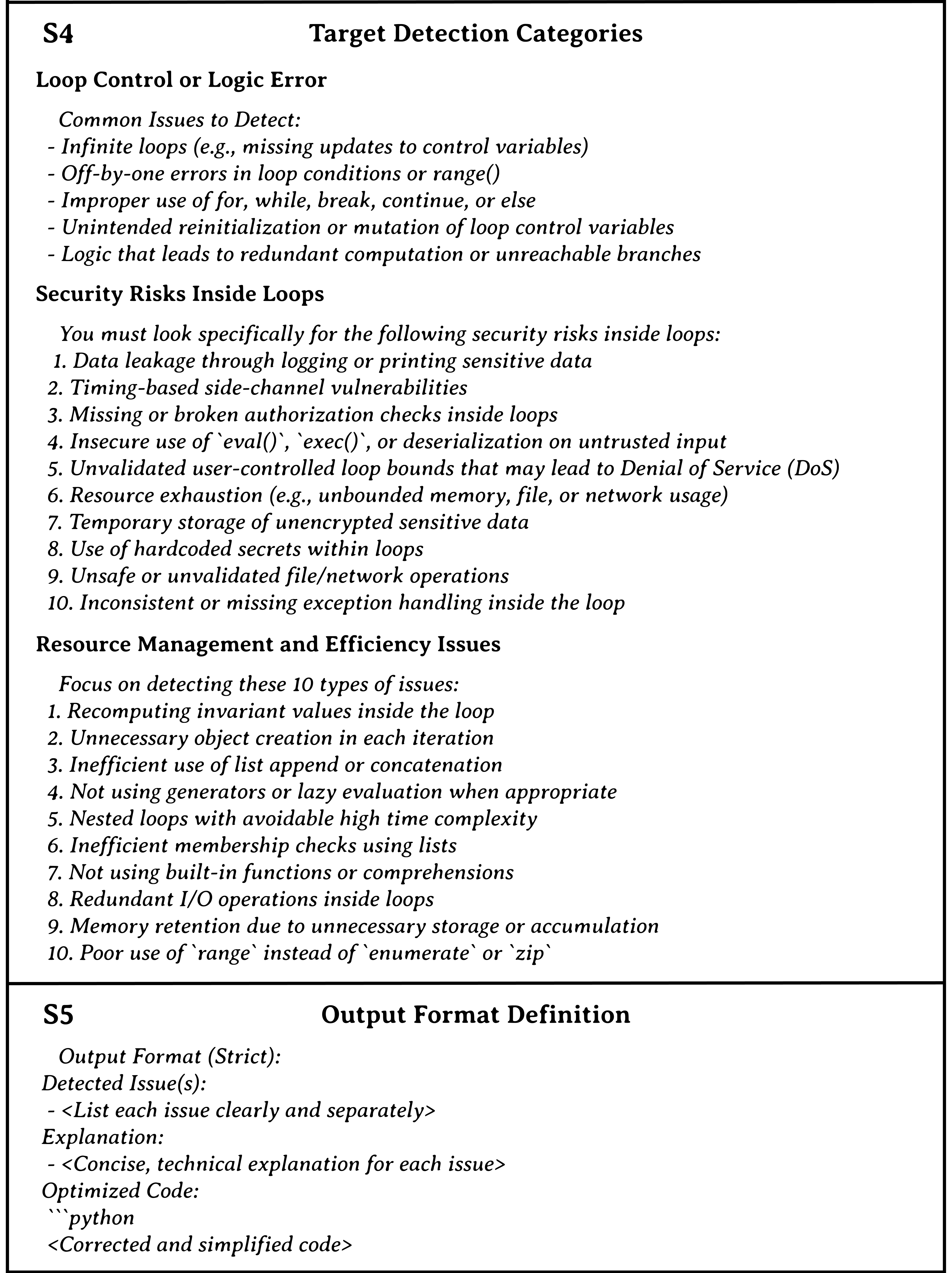}
    \caption{Block 4 \& 5 of the structure of the system prompt to detect loop vulnerabilities}
    \label{fig:b45}
\end{figure}

Figure \ref{fig:b123} and \ref{fig:b45} presents a structured system prompt designed for use with LLaMA and Phi. The prompt guides the model in detecting Python loop-related issues, including logic errors, security risks, and efficiency problems.

It is divided into five sections (S1–S5), each addressing a specific function: system identity and capabilities, core-functional responsibilities, constraints and guardrails, target detection categories and output format definition. The design included key safeguarding features such as language-specific awareness, code-aware grounding, version sensitivity, and hallucination prevention. In S2 and S4, only the required loop vulnerability category should be retained for detection. All other vulnerability category information must be removed or filtered out before use in these sections.  Hence, precision, relevance, and safety are ensured, which is crucial for LLM-assisted code analysis. Also, the user prompt contains each code block that was passed for analysis.

\section{Validation and Evaluation}

Three standard evaluation metrics, precision, recall, and F1-score, were utilized in assessing LLaMA and Phi's performance for loop vulnerability detection. The metrics are defined mathematically as follows: 
\begin{itemize}
\item \textbf{Precision:} Precision is a measure of the proportion of true positives (TP) against all predicted positives (TP + FP). It indicates how often the model's positive predictions are correct \cite{ref_proc10}.
\[
\text{Precision} = \frac{TP}{TP + FP}
\]
\item \textbf{Recall:} Recall is a measure of the proportion of true positives that were correctly identified when compared to all actual positives (TP + FN) \cite{ref_proc10}.
\[
\text{Recall} = \frac{TP}{TP + FN}
\]

\item \textbf{F1-score:} The F1-score is the harmonic mean of precision and recall. It balances both metrics and is mainly useful when there is an uneven class distribution or when false positives and negatives are relevant \cite{ref_proc10}.
\[
\text{F1-score} = 2 \times \frac{\text{Precision} \times \text{Recall}}{\text{Precision} + \text{Recall}}
\]

\end{itemize}
These metrics help present a comprehensive overview of the performance of the model. The results of the three loops vulnerability type for each model (LLaMA \& Phi) can be seen in Table \ref{tab:validation}. The result indicates that Phi consistently outperforms LLaMA in precision, recall, and F1-score in all three categories, 0.90 for loop control and logic errors, as well as security risks, and  0.95 for resource management issues. This highlights Phi's stronger precision and recall in all categories. LLAMA performed competitively but with marginally reduced precision and recall when compared to Phi.

\begin{table}[]
\caption{Validation Results of LLaMA and Phi}
\label{tab:validation}
\begin{tabular}{|lllllllllllll|}
\hline
\multicolumn{1}{|l|}{}                 & \multicolumn{6}{c|}{\textbf{LLaMA}}                                                                                                                                                                                        & \multicolumn{6}{c|}{\textbf{Phi}}                                                                                                                                                                     \\ \hline
\multicolumn{1}{|l|}{}                 & \multicolumn{1}{l|}{TP}        & \multicolumn{1}{l|}{FP}        & \multicolumn{1}{l|}{FN}        & \multicolumn{1}{l|}{\textbf{Precision}} & \multicolumn{1}{l|}{\textbf{Recall}} & \multicolumn{1}{l|}{\textbf{F1-score}} & \multicolumn{1}{l|}{TP}        & \multicolumn{1}{l|}{FP}        & \multicolumn{1}{l|}{FN}        & \multicolumn{1}{l|}{\textbf{Precision}} & \multicolumn{1}{l|}{\textbf{Recall}} & \textbf{F1-score} \\ \hline
\multicolumn{13}{|c|}{\textbf{Loop Control and Logic Errors}}                                                                                                                                                                                                                                                                                                                                                                                                               \\ \hline
\multicolumn{1}{|l|}{Code 1}           & \multicolumn{1}{l|}{6}         & \multicolumn{1}{l|}{2}         & \multicolumn{1}{l|}{1}         & \multicolumn{1}{l|}{0.75}               & \multicolumn{1}{l|}{0.86}            & \multicolumn{1}{l|}{0.80}              & \multicolumn{1}{l|}{6}         & \multicolumn{1}{l|}{1}         & \multicolumn{1}{l|}{1}         & \multicolumn{1}{l|}{0.86}               & \multicolumn{1}{l|}{0.86}            & 0.86              \\ \hline
\multicolumn{1}{|l|}{Code 2}           & \multicolumn{1}{l|}{7}         & \multicolumn{1}{l|}{1}         & \multicolumn{1}{l|}{1}         & \multicolumn{1}{l|}{0.75}               & \multicolumn{1}{l|}{0.86}            & \multicolumn{1}{l|}{0.80}              & \multicolumn{1}{l|}{8}         & \multicolumn{1}{l|}{0}         & \multicolumn{1}{l|}{0}         & \multicolumn{1}{l|}{0.86}               & \multicolumn{1}{l|}{0.86}            & 0.86              \\ \hline
\multicolumn{1}{|l|}{Code 3}           & \multicolumn{1}{l|}{6}         & \multicolumn{1}{l|}{1}         & \multicolumn{1}{l|}{1}         & \multicolumn{1}{l|}{0.88}               & \multicolumn{1}{l|}{0.88}            & \multicolumn{1}{l|}{0.88}              & \multicolumn{1}{l|}{7}         & \multicolumn{1}{l|}{1}         & \multicolumn{1}{l|}{0}         & \multicolumn{1}{l|}{1}                  & \multicolumn{1}{l|}{1}               & 1                 \\ \hline
\multicolumn{1}{|l|}{\textbf{Average}} & \multicolumn{1}{l|}{\textbf{}} & \multicolumn{1}{l|}{\textbf{}} & \multicolumn{1}{l|}{\textbf{}} & \multicolumn{1}{l|}{\textbf{0.79}}      & \multicolumn{1}{l|}{\textbf{0.86}}   & \multicolumn{1}{l|}{\textbf{0.83}}     & \multicolumn{1}{l|}{\textbf{}} & \multicolumn{1}{l|}{\textbf{}} & \multicolumn{1}{l|}{\textbf{}} & \multicolumn{1}{l|}{\textbf{0.90}}      & \multicolumn{1}{l|}{\textbf{0.90}}   & \textbf{0.90}     \\ \hline
\multicolumn{13}{|c|}{\textbf{Security Risks Inside Loops}}                                                                                                                                                                                                                                                                                                                                                                                                                 \\ \hline
\multicolumn{1}{|l|}{Code 4}           & \multicolumn{1}{l|}{13}        & \multicolumn{1}{l|}{3}         & \multicolumn{1}{l|}{2}         & \multicolumn{1}{l|}{0.81}               & \multicolumn{1}{l|}{0.87}            & \multicolumn{1}{l|}{0.84}              & \multicolumn{1}{l|}{14}        & \multicolumn{1}{l|}{1}         & \multicolumn{1}{l|}{1}         & \multicolumn{1}{l|}{0.93}               & \multicolumn{1}{l|}{0.93}            & 0.93              \\ \hline
\multicolumn{1}{|l|}{Code 5}           & \multicolumn{1}{l|}{11}        & \multicolumn{1}{l|}{1}         & \multicolumn{1}{l|}{1}         & \multicolumn{1}{l|}{0.81}               & \multicolumn{1}{l|}{0.87}            & \multicolumn{1}{l|}{0.84}              & \multicolumn{1}{l|}{10}        & \multicolumn{1}{l|}{2}         & \multicolumn{1}{l|}{2}         & \multicolumn{1}{l|}{0.93}               & \multicolumn{1}{l|}{0.93}            & 0.93              \\ \hline
\multicolumn{1}{|l|}{Code 6}           & \multicolumn{1}{l|}{9}         & \multicolumn{1}{l|}{1}         & \multicolumn{1}{l|}{2}         & \multicolumn{1}{l|}{0.92}               & \multicolumn{1}{l|}{0.92}            & \multicolumn{1}{l|}{0.92}              & \multicolumn{1}{l|}{11}        & \multicolumn{1}{l|}{0}         & \multicolumn{1}{l|}{0}         & \multicolumn{1}{l|}{0.83}               & \multicolumn{1}{l|}{0.83}            & 0.83              \\ \hline
\multicolumn{1}{|l|}{\textbf{Average}} & \multicolumn{1}{l|}{\textbf{}} & \multicolumn{1}{l|}{\textbf{}} & \multicolumn{1}{l|}{\textbf{}} & \multicolumn{1}{l|}{\textbf{0.85}}      & \multicolumn{1}{l|}{\textbf{0.88}}   & \multicolumn{1}{l|}{\textbf{0.86}}     & \multicolumn{1}{l|}{\textbf{}} & \multicolumn{1}{l|}{\textbf{}} & \multicolumn{1}{l|}{\textbf{}} & \multicolumn{1}{l|}{\textbf{0.90}}      & \multicolumn{1}{l|}{\textbf{0.90}}   & \textbf{0.90}     \\ \hline
\multicolumn{13}{|c|}{\textbf{Resource Management and Efficiency Issues}}                                                                                                                                                                                                                                                                                                                                                                                                   \\ \hline
\multicolumn{1}{|l|}{Code 7}           & \multicolumn{1}{l|}{14}        & \multicolumn{1}{l|}{1}         & \multicolumn{1}{l|}{2}         & \multicolumn{1}{l|}{0.93}               & \multicolumn{1}{l|}{0.88}            & \multicolumn{1}{l|}{0.90}              & \multicolumn{1}{l|}{15}        & \multicolumn{1}{l|}{0}         & \multicolumn{1}{l|}{1}         & \multicolumn{1}{l|}{1}                  & \multicolumn{1}{l|}{0.94}            & 0.97              \\ \hline
\multicolumn{1}{|l|}{Code 8}           & \multicolumn{1}{l|}{12}        & \multicolumn{1}{l|}{2}         & \multicolumn{1}{l|}{1}         & \multicolumn{1}{l|}{0.93}               & \multicolumn{1}{l|}{0.88}            & \multicolumn{1}{l|}{0.90}              & \multicolumn{1}{l|}{12}        & \multicolumn{1}{l|}{1}         & \multicolumn{1}{l|}{1}         & \multicolumn{1}{l|}{1}                  & \multicolumn{1}{l|}{0.94}            & 0.97              \\ \hline
\multicolumn{1}{|l|}{Code 9}           & \multicolumn{1}{l|}{10}        & \multicolumn{1}{l|}{0}         & \multicolumn{1}{l|}{1}         & \multicolumn{1}{l|}{0.86}               & \multicolumn{1}{l|}{0.92}            & \multicolumn{1}{l|}{0.89}              & \multicolumn{1}{l|}{11}        & \multicolumn{1}{l|}{0}         & \multicolumn{1}{l|}{0}         & \multicolumn{1}{l|}{0.92}               & \multicolumn{1}{l|}{0.92}            & 0.92              \\ \hline
\multicolumn{1}{|l|}{\textbf{Average}} & \multicolumn{1}{l|}{\textbf{}} & \multicolumn{1}{l|}{\textbf{}} & \multicolumn{1}{l|}{\textbf{}} & \multicolumn{1}{l|}{\textbf{0.91}}      & \multicolumn{1}{l|}{\textbf{0.89}}   & \multicolumn{1}{l|}{\textbf{0.90}}     & \multicolumn{1}{l|}{\textbf{}} & \multicolumn{1}{l|}{\textbf{}} & \multicolumn{1}{l|}{\textbf{}} & \multicolumn{1}{l|}{\textbf{0.97}}      & \multicolumn{1}{l|}{\textbf{0.93}}   & \textbf{0.95}     \\ \hline
\end{tabular}
\end{table}

\vspace{-10pt}

\section{Discussion and Conclusion}
This study developed a framework for detecting loop-related vulnerabilities in Python code using local LLMs. Traditionally, static and dynamic analysis tools often miss semantically complex loop errors because they rely on syntactic pattern matching, extensive test data, and execution environment. Our framework uses local LLMs, LLaMA and Phi, to detect loop-related vulnerabilities. It focuses on deeper semantic understanding and targets three types of vulnerabilities: loop control and logic errors, security risks inside loops, and resource management and efficiency issues. 

The results indicate that Phi performed better than LLaMA across all three vulnerability categories, achieving higher F1-scores, especially in detecting efficiency-related issues. Our methodology involved manual annotation to create a validated ground-truth, followed by iterative prompt engineering to refine model behavior. We also adopted a three-phase validation process to compare model outputs against the manually verified baseline ground-truth. The system prompt assigned the model a code optimization reviewer role to detect loop vulnerabilities, while the user prompts provided the code block to investigate. This layered prompt structure helped reduce hallucination, narrow task scope, and improve reproducibility. 

The framework has limitations; the approach used in this study cannot detect concurrency and synchronization issues. Detecting often requires special techniques such as temporal reasoning and dynamic code analysis. However, these techniques are not suitable for the prompt-based approach used in this study. Future work could focus on introducing a framework that can detect concurrency issues as well and support additional programming languages. Larger local but code-specific models like CodeBERT or CodeT5 could be compared as well for accuracy. Also, integration into real-time development environments and Integrated Development Environments (IDEs) will make this approach more practical.

\end{document}